# TWO-POINT ANGULAR CORRELATION FUNCTION FOR THE GREEN BANK 4.85 GHZ SKY SURVEY

Brian L. Kooiman, Jack O. Burns, and Anatoly A. Klypin

Department of Astronomy, Box 30001/Dept. 4500,
New Mexico State University, Las Cruces, NM 88003

## ABSTRACT

This paper presents an angular correlation analysis of the Green Bank 4.85 GHz radio catalog (Gregory & Condon 1991) of 54,579 sources (S $\gtrsim$ 25 mJy). The Green Bank catalog is found to be complete to S $\geq$ 35 mJy over $20° \leq \delta < 74°$, $0^h \leq \alpha < 24^h$, and Galactic latitude $|b| \geq 10°$. The 2-point angular correlation function shows evidence for the clustering of radio sources, with a power-law distribution consistent with a slope $\gamma = -0.8$. This may well provide the *first* detection of an angular correlation in a large area, complete deep radio survey.

*Subject headings:* radio sources: clustering - cosmology: large scale structure of the Universe

# 1 Introduction

The angular and spatial correlation functions have been commonly used statistics for analyzing clustering in the galaxy distribution. Since few 3-dimensional surveys exist, the angular correlation function has been employed more extensively, as this only requires the 2-dimensional projection of galaxies. Although not as powerful as the spatial correlation function, the 2-point angular correlation $w(\theta)$ can be related to the 2-point spatial correlation function $\xi(r)$ (Limber 1953), gaining valuable insights on the clustering of galaxies.

There has been a great deal of work done in the optical regime with the 2-point angular correlation function. Recent results include the the Automated Plate Machine (APM) survey by Maddox et al. (1990), where $w(\theta)$ is estimated for $\gtrsim$ 2 million southern hemisphere galaxies complete to $b_J = 20.5$. Similar to the Lick results (Groth & Peebles 1977), the APM correlation function also shows evidence of the power-law behavior $w(\theta) \propto \theta^{1-\gamma}$, with $\gamma \approx 1.7$ on small scales, and a break on larger scales. These results are consistent with estimates given by Picard (1991; Palomar Observatory Sky Survey II), and Collins, Nichol, & Lumsden (1992; Edinburgh-Durham Southern Galaxy Catalogue).

Radio surveys provide another very different sample for 2-point studies. Many of the optical surveys tend to cover limited areas of several arcminutes$^2$ or several degrees$^2$, sampling clustering on small scales (notable exceptions being plate surveys by Groth & Peebles (1977), Maddox et al. (1990), Picard (1991), and Collins et al. (1992)). Radio surveys, however, tend to cover much larger areas and sample far greater volumes - albeit with sparser coverage. Consequently, radio surveys allow an investigation of structure on greater physical scales than most optical studies. Furthermore, the virtual absence of obscuration and confusion is also unique to modern radio surveys.

The use of radio surveys for investigating large scale structure is not new. By studying the angular distribution of $\sim 8000$ sources at 408 MHz, Webster (1977) found an upper limit of 3 % for the variation of the number of radio sources in a randomly placed 1 Gpc cube. The general consensus which emerged was that the distribution of radio sources is isotropic. If any large scale structure was present, then such a signal would be washed out by the broad luminosity function of extragalactic radio sources; since any flux density limited sample has objects from a broad range of redshifts, the resultant projection of all these sources onto the plane of the sky would dilute any evidence of large scale structure.

Recently, Shaver & Pierre (1989) presented evidence that the distribution of strong extragalactic radio sources may not be isotropic. Their analysis of the 408 MHz Molonglo Reference Catalog of Radio Sources (Large et al. 1981) indicates clustering of radio sources near the supergalactic plane for z < 0.02. Peacock & Nicholson (1991) gave an estimate of the power spectrum and the spatial correlation function for $\sim 300$ 1.4 GHz radio sources with $0.01 < z < 0.1$. Their results are consistent with $\xi(r) \simeq (r/11h^{-1}\mathrm{Mpc})^{-1.8}$.

With the release of new radio surveys at 1.4 GHz (Condon & Broderick 1985, 1986) and 4.85 GHz (Condon, Broderick, & Seielstad 1989, (hereafter CBS); Griffith 1993), the quality and quantity of the datasets are greatly improved. As a result, better constraints can now be placed on large scale structure. Griffith (1993) has presented a preliminary $w(\theta)$ estimate for 4.51 steradians of the Parkes-MIT-NRAO (PMN) southern hemisphere survey. He finds no evidence of clustering for $60 < S < 125$ mJy and $125 \leq S \leq 1000$ mJy. However, by splitting his sample into two flux density intervals, the number of sources in each was reduced to 7300 and 3000 sources, respectively, making the statistical noise higher. Additionally, he reduced the amplitude of the potential signal in the $60 < S < 125$ mJy sample by excluding the highest flux density sources. Our analysis of $w(\theta)$ for the Gregory & Condon (1991) catalog in Section 3 does show a signal for the flux density limit $S \geq 60$ mJy since our extended flux density range includes more sources and the strongest sources, wherein the clustering should be most prevalent. For all sources with $S \geq 60$ mJy, we find a signal in $w(\theta)$ at a level of $\approx 3\sigma$.

The CBS northern hemisphere survey at 4.85 GHz provided our testground for examining $w(\theta)$ of radio sources. The catalog listing of 54,579 sources by Gregory & Condon (hereafter 87GB), covers 6.0 steradians down to a sensitivity of $\sim 25$ mJy. Consequently, unlike many small area or small source number catalogs, the 87GB catalog covers a much larger region of the sky and contains far more sources. For comparison, the 87GB catalog has over 10 times the number of sources as the 4C catalog (Pilkington & Scott 1965; Gower et al. 1967) and well over 100 times the number of sources in the 3C catalog (Bennett 1962). Such a large number of sources should provide an excellent statistical base with which to work, helping confine the scale of hierarchical galaxy clustering more stringently.

We begin this paper by establishing the completeness of the 87GB catalog in Section 2. Next, we select an estimator for the 2-point angular correlation function and present our estimates for $w(\theta)$ in Section 3. Then, we contrast theoretical implications with our 87GB results in Section 4. Finally, we present our conclusions in Section 5.

## 2 THE 87GB CATALOG

### 2.1 Overview

The Green Bank 4.85 GHz Northern Sky Survey (CBS) used the NRAO 300 ft (91 m) telescope to survey the sky between $0°$ and $+75°$ declination. The survey consists of transit elevation scans made by a seven-beam receiver which produces seven parallel tracks with a separation of $\sim 1$ FWHM beamwidth. The scans were made by driving the telescope alternately north and south at its maximum slew rate of $10°$ min$^{-1}$. The data processing that might be important for our correlation analysis (e.g., corrections for gain changes or subtraction of spillover baselines) was done for eight strips of constant declination centered on $\delta = 0°, +10°, ..., +70°$. Then the data were gridded into 285 FITS images of size 1024 x 1024 with 40" pixels. The resultant images were reduced with the NRAO AIPS software to yield sky maps covering 6.0 steradians with a resolution of $3.''7 \times 3.''3$ and a RMS noise limit of $\sim 5$ mJy (also the confusion limit). Initially, 288 maps were constructed, but three maps were rejected due to strong solar interference. Additional limited areas of solar interference and inadequate data were also excised to leave the resultant 6.0 steradian survey. In Figure 1 we show the 87GB catalog of 54,579 sources with angular sizes less than $10.''5$ and S $\gtrsim 25$ mJy.

## 2.2 Completeness

In order to assess the completeness of this catalog, we initially constructed a series of Aitoff equal area projection plots in 5 mJy flux density intervals. Becker, White, & Edwards (1991, hereafter BWE) have compiled their own catalog of 53,522 sources based on the the CBS survey, and claim a completeness level of 25 mJy for $\delta > 20°$ and $|b| \geq 5°$. Because of the well known decrease in sensitivity of the 300 ft Green Bank telescope at the lowest declinations, we also opted to use the same declination limit as BWE.

However, even though we retained the declination limits of BWE, we excluded all sources within 10° of the Galactic plane, rather than their choice of 5°. This exclusion should provide a more conservative means of removing radio sources associated with the Galactic population – consistent with our cautious approach to source selection. Additionally, Griffith (1993) found that for $S \geq 60$ mJy, the over-density of 4.85 GHz Galactic sources drops to 1% at $|b| = 10°$. Hence, beyond our cutoff, the Galactic population is expected to comprise a negligible fraction of the radio source distribution.

Figures 2a and 2b show the 87GB catalog from $20° \leq \delta < 75°$, $0^h \leq \alpha < 24^h$, and $|b| \geq 10°$, and over two flux density ranges: $25 \leq S < 30$ mJy and $35 \leq S < 40$ mJy. An initial examination of the 25-30 mJy sample of the 87GB catalog shows this flux density interval is far from complete. Not only is there a paucity of sources near $\alpha \simeq 13^h$, but the lowest declinations also suffer from incompleteness. Conversely, the 35-40 mJy 87GB sample shows no visible indication of non-homogeneity. Thus, whereas the distribution of all sources appeared fairly uniform, a closer inspection in discrete flux density intervals shows a definite level of incompleteness at the lowest flux density levels. Consequently, our initial conclusion was that the 87GB catalog appears to be complete above a flux density of 35 mJy and over $20 \leq \delta < 75°$, and $|b| \geq 10°$.

To further investigate the completeness of the 87GB catalog, we calculated the surface number density of sources for a series of limiting flux density values. This entailed tallying the number of sources in 1° strips of declination, and plotting the resultant number per square degree. Error bars were determined by $\sqrt{N_i}/A_i$ error bars, where $N_i$ is the number of sources in bin $i$, and $A_i$ is the effective area of bin $i$. Figure 3 shows the surface number density as a function of declination for a flux density cutoff of 35 mJy. The final declination bin ($74° \leq \delta < 75°$) is particularly aberrant. Condon (1992) suggests that the enhanced density in this highest $\delta$ bin might be an artifact of the baseline subtraction, which could be less accurate at the highest declinations. However, for the declination range $20° - 74°$, the surface number density appears constant within the expected errors. Consequently, we ignored all sources with $\delta \geq 74°$, and limited further investigation to the declination range $20° - 74°$.

There is some indication of a sine wave in the surface number density plot, particularly at the lowest declinations. In order to make a statistical judgement about irregularities in the surface number density, we constructed a series of periodograms. Figure 4 shows the power spectra for a flux density cutoff of 35 mJy. The angular frequency is defined as $\Omega = 2\pi/T$, where $T$ is the period of the wave in degrees, and the power is a measure of the probability that such an angular frequency is real. The level of power fluctuations for randomly distributed points is indicated by the dashed curve, corresponding to a 95% confidence level. A total of 50 random samples was used to determine the periodogram confidence level. (The drop in power for the smallest angular frequencies is the result of subtracting an average surface number density from each realization – data and random.) The tendency for the 35 mJy 87GB periodogram to fall below the two sigma limit indicates that this sample contains no statistical periodicities. Furthermore, the frequencies below $\sim 0.1$ correspond to periodicities longer than the $54°$ declination range of the catalog, and thus are sensitive to linear trends in the data.

We caution against using sources below 35 mJy for statistical analyses. Not only do the $25-30$ mJy Aitoff equal area projection plots show serious incompleteness, but the $S \geq 25$ mJy surface number density strips also show a disturbing gradient and apparent $\sim 10°$ periodicity. Whereas the gradient may in part be related to the poorer sensitivity of the 300 ft Green Bank telescope at the lowest declinations, we can only conjecture that any possible periodicity arises from the data acquisition method, since no similar effect is seen in Galactic surface number density plots. Hence, we discourage using the $S < 35$ mJy component of the 87GB catalog for statistical studies.

To summarize, we have employed three distinct means to establish the completeness of the 87GB catalog. Based on Aitoff projection plots, surface number density strips, and a periodogram analysis, we have chosen a completeness limit of $S \geq 35$ mJy over $20° \leq \delta < 74°$, $0^h \leq \alpha < 24^h$, and $|b| \geq 10°$. Figure 5 shows this complete 87GB catalog containing 21,490 sources. Yet another confirmation of completeness follows from the analysis of $w(\theta)$ on very large scales (up to $140°$) in Section 3. Depending on the particular correlation function estimator, the amplitude of $w(\theta)$ on large scales is $\lesssim 10^{-2}$ for the poorest estimator and $\lesssim 10^{-3}$ for the most accurate estimator, signaling the absence of a global offset from a gradient in the sample. Consequently, we believe our conservative choice of flux density cutoff and boundaries should allow a viable $w(\theta)$ analysis.

## 3 2-POINT ANGULAR CORRELATION FUNCTION

The 2-point angular correlation function $w(\theta)$ provides a measure of galaxy density excess over that expected for a random distribution of sources. Our interest in estimating $w(\theta)$ for the 87GB catalog stems from the catalog's unique nature. The CBS survey presents an opportunity to investigate the clustering of radio sources not only on larger scales, but out to far greater depths as well; whereas an optical survey typically does not extend to high $z$, a radio survey like the CBS contains sources beyond $z = 1$. Additionally, the large number of sources and homogeneous coverage of a large fraction of the sky (6.0 steradians) provide an excellent statistical database. Hence, we believe the 2-point analysis of the 87GB catalog should prove insightful.

Because the amplitude of $w(\theta)$ is very small and could be influenced by uncertainty of the particular correlation estimator or a small residual observational contamination, we analyzed the data in a variety of ways, examining three estimators, four different sky subsamples, and four flux density cutoffs. Two traditional estimators were investigated: $w(\theta) = DD(\theta)/RR(\theta) - 1$, and $w(\theta) = DD(\theta)/DR(\theta) - 1$, where $DD(\theta)$ is the number of data-data pairs, $RR(\theta)$ is the number of random-random pairs, and $DR(\theta)$ is the number of data-random pairs – all for angular separation $\theta \pm \delta\theta$. Recently, Hamilton (1993a) proposed an additional estimator, which is claimed to be more reliable on larger scales. This angular correlation function is defined as

$$w(\theta) = \frac{DD(\theta) \cdot RR(\theta)}{(DR(\theta))^2} - 1. \tag{1}$$

Hamilton states that whereas the uncertainty of the previous estimators is limited by the uncertainty in the mean density of the sample, his formulation is limited by pair counts rather than the mean density. Thus, his estimator should prove more reliable, particularly on scales where the amplitude of the angular correlation function is small.

### 3.1 Comparison of 2-Point Angular Correlation Estimators and Associated Errors

To estimate $w(\theta)$, we constructed 100 random catalogs with the same number of points as the number of sources in the particular subsample of the 87GB catalog. The number of random catalogs was large enough to ensure that the error in the mean number of random pairs was much smaller than the expected Poissonian noise for the same size data catalog. Figure 6 shows the three 2-point angular correlation functions for the complete 87GB catalog. The open triangles correspond to the DD/RR formulation, the filled squares to the DD/DR formulation, and the filled circles to the (DD·RR)/(DR)$^2$ formulation. Since Poissonian error bars are used, the error bars of the first two formulations are identical to those plotted for Hamilton's formulation. (The angular separation in these correlations only extends to 140° since this is the greatest separation for any two points in our $\delta = +20°$ limited sample.) At large angular scales, the first two formulations have rather large non-zero oscillations, whereas the oscillations in Hamilton's formulation remain roughly an order of magnitude smaller. Thus, it appears the different formulations treat boundary regions quite differently, and that Hamilton's formulation is indeed the most robust at large angular scales.

Although it is valuable to know the reliability of $w(\theta)$ on large scales to help distinguish formulation irregularities from the effects of catalog incompleteness or gradients, we are most concerned with $w(\theta)$ on small angular scales, where a non-zero amplitude may be physically significant. Figure 7 shows these same formulations, but only out to 5°. All three have similar amplitudes, although Hamilton's formulation often is slightly larger. However, even though the three formulations are very similar on the scales of most interest, we have opted to use Hamilton's formulation in the analysis below. The removal of the mean density uncertainty in this formulation apparently results in a general robustness on large scales, as well as a close agreement with the other formulations at the smallest scales. Thus, we believe that the (DD·RR)/(DR)$^2$ formulation provides the truest correlation.

Although we cannot ignore that our correlation pairs come from the same set of individual galaxies, and thus, are not totally independent of one another, our choice of Poissonian error bars is not without forethought. First, one must recognize that the problem of inter-related pairs is most severe with small samples. However, with many thousands of sources in our catalogs, the effects of small sample sizes should not be troublesome. And second, we acknowledge the claims made by Landy & Szalay (1993) and Hamilton (1993a) that the variance of some estimators can be larger than Poissonian. Consequently, the DD/RR and DD/DR error bars should be treated as lower limits. However, Hamilton's estimator has been shown to be nearly Poissonian in variance (Landy & Szalay 1993), with the additional bias being the "discreteness correction" (Hamilton 1993b) - equivalent to using N(N-1) in place of N$^2$. But for our catalogs with large N, this bias should be inconsequential.

## 3.2  2-Point Results Over Various Regions

Having settled on a formulation for $w(\theta)$, we proceeded to investigate the consistency of our results - namely, if the correlations would vary over different regions of the sky or remain region independent. To do this, we examined several different samples of the 87GB catalog. The limits of these samples are shown in Table 1.

Figure 8 shows $w(\theta)$ estimated for each of the regions in Table 1. As expected, the error bars increase as the number of sources in each successive region is reduced. Note that the same general features are seen in each region: the absent signal for $\theta < 12'$, a $\sim 3\sigma$ anti-correlation in the fourth bin (48'-60'), and a similar shape for each of the 2-point functions. The "absent" signal below 12' is actually a highly negative correlation - typically with amplitude near -0.1. However, the paucity of 87GB sources at the smallest separations stems from confusion caused by the 3.7 ' $\times$ 3.3 beam. Thus, we ignored this signal, and scaled the plots to an amplitude range wherein the results were physically meaningful.

Conversely, the cause of the deviant fourth bin is unknown. Not only does this feature persist when the bin size is altered, and when a new set of 100 random catalogs is substituted in the RR and DR calculations, but it also remains when the DD/RR and DD/DR formulations of $w(\theta)$ are used. Consequently, we believe the aberrant bin is not a statistical fluctuation unique to our analysis, but is somehow inherent to the data. Hence, these results should prove reliable, providing a rigorous representation of the 87GB correlation function.

## 3.3  Varying Flux Density Cutoffs

Four different flux density limits were chosen, with cutoffs at 35, 45, 60, and 300 mJy, and with the coverage region identical to Figure 5. The 2-point angular correlation function for these cutoffs is shown in Figure 9 (note the scale change for Figure 9d). Whereas the 35 mJy sample of 21,490 sources, the 45 mJy sample of 15,398 sources, and the 60 mJy sample of 10,554 sources are rather similar, the 300 mJy sample of 1,067 sources appears quite different. Not only does the $\theta < 12'$ signal have a *positive* amplitude in this smallest sample, but the next four correlations toggle negative and positive. In comparison, the 35, 45, and 60 mJy samples have only one negative correlation in bins $1 - 7$ ($0.2° - 1.6°$). Here, it is reassuring to see that our results do not fluctuate wildly when the flux density cutoff is changed from 35 to 45 to 60 mJy.

When producing the 87GB catalog, Gregory & Condon (1991) made corrections for a number of effects, which change with $\delta$ but do not depend on $\alpha$ (e.g., corrections for gain changes or subtraction of spillover baselines). This potentially could produce spurious signals in the correlation function if some residual mismatch remained in the data. In order to check for a possible systematical effect, we calculated the angular correlation function for each of original 10° strips and after that averaged the results. This procedure removes any correlation due to possible mismatch between strips of constant declination. Unfortunately, it also reduces the number of pairs, thus reducing the signal-to-noise level. Triangles in Figure 9a present the correlation function for this case. We find that within the noise limits the correlation function for the summed declination strips agrees with that found for the all-catalog correlation function.

A closer examination of the 35, 45, and 60 mJy correlation functions reveals two interesting trends. First, the highly deviant bin ($0.8° - 1.0°$) becomes more negative with increasing flux density cutoff. This provides some degree of assurance that this signal, although real, is more accentuated at higher flux densities. And second, the positive amplitude of the correlation function on small scales becomes weaker with decreasing flux density. Thus, the 60 mJy sample shows the strongest evidence for radio source clustering at small angular separation.

Additionally, a comparison of all four samples shows a significant change in the associated error bars. Here, the 300 mJy sample should provide a rough comparison to the 3C catalog, although the selection of sources in the CBS survey at 4.85 GHz is different from that of the 3C at 178 MHz. Notice that the associated error bars of the 300 mJy sample are far too large to place tight constraints on extragalactic clustering. However, the 35, 45, and 60 mJy samples highlight the utility of the 87GB catalog, wherein the large number of sources reduces the error bars, and facilitates a more fruitful investigation of large scale structure. Section 4 will examine the clustering implications of $w(\theta)$ for the 87GB 35 mJy catalog more closely.

With the exception of the 300 mJy sample, all the subsamples indicate the presence of a cosmological signal on small angular scales. To better assess the statistical significance of this signal, we examined an "integrated" correlation function, wherein the size of the first bin was successively enlarged from 0.2° to 4.6°. (We excluded the angular separations $0.0° \leq \theta < 0.2°$ and $0.8° \leq \theta < 1.0°$, since they were not indicative of the collective signal.) Here, the number of data pairs in the expanding bin was compared with the Poissonian expectation for the same size angular bin. Hence, this test is a cumulative analysis which quantifies the significance of $w(\theta)$ for all the data, and not just an individual angular separation. Figure 10 shows the deviation of pair counts from the expected value of a random distribution, where the deviation is estimated as $\sigma = (DD - RR)/(RR)^{1/2}$. The solid line corresponds to the 35 mJy sample, the dashed line to the 45 mJy sample, and the dotted line to the 60 mJy sample. Note that the points in this plot are not statistically independent; the maximum deviation indicates a statistical significance level for which we can reject the hypothesis that the observationed correlation function arises from a random distribution. Hence, the $\sim 3\sigma$ deviation in the 35 mJy sample indicates that the non-zero amplitude of $w(\theta)$ is statistically significant.

### 3.4 Power-law Approximation

The observed signal was fit to a power law of the form $w(\theta) = a\theta^\gamma$ over the range $0.3° - 1.9°$ (9 bins) with the weight of each point proportional to the number of pairs in each bin. The solid line in Figure 11 shows the fit to all nine data points, and the dashed line shows a forced fit of slope -0.8. Although the general amplitude of the forced fit is too small, the slope agrees rather nicely with the data (if one dismisses the negative bin). This implies the 87GB survey may be somewhat consistent with the canonical power-law index of -0.8.

For comparison, a fit to just the eight positive points is shown as the dotted line in Figure 11. It is readily apparent that removing the negative point flattens the slope significantly, from -1.27 to -0.76; when the fourth of nine logarithmic bins is negative, the amplitude at the greatest separations is pulled down, resulting in a steeper slope. Now, the inclusion of this final fit should not be interpreted as grounds to invalidate the negative bin. Rather, we constructed this fit to gain some insight on how closely the positive correlation bins follow a power-law. The proximity of this fit's slope to -0.8 was pleasantly reassuring. Amplitudes, exponents, and chi-squared values for all three fits appear in Table 2.

## 4 THEORETICAL 2-POINT ANGULAR CORRELATION FUNCTION

Having calculated $w(\theta)$ for the 87GB catalog, we can now investigate what these results can tell us about the clustering of radio sources at 4.85 GHz. Because of uncertainties in the evolution of both the luminosity function of radio sources and the correlation function $\xi(r,z)$, it is difficult to make any definite predictions. We start with estimates of $w(\theta)$ assuming that the local Radio Luminosity Function does not change with redshift, and that $\xi(r,z)$ has the observed shape at $z = 0$ and obeys self-similar scaling at higher $z$. The model is too naive to be realistic, but it provides us with a reference point. We discuss effects of possible changes in $\xi(r,z)$ and RLF at the end of this section

We decided to use a power-law model for the spatial distribution of radio sources which is of the form $\xi(r) = (r/r_0)^{-\gamma}(1+z)^{-(3+\epsilon)}$, where $r$ is the proper spatial separation, $r_0$ is the correlation scale length at $z = 0$, and $\gamma$ is the power-law index equal to 1.8. The final term models clustering evolution with redshift. If $\epsilon = -1.2$, then the correlation function is constant in comoving coordinates. We opted to use $\epsilon = 0$, which corresponds to classical self-similar clustering. Altering the value of $\epsilon$ only resulted in minimal changes to the resultant $w(\theta)$. Setting $\gamma = 1.8$ for the spatial distribution of sources in our model, we used the relativistic version of Limber's equation (Limber 1953) as it appears in Peebles (1980) to calculate $w(\theta)$ from $\xi(r)$. Values of $H_0 = 50$ km s$^{-1}$Mpc$^{-1}$, and $\Omega_0 = 1$ were assumed in all calculations.

Since our goal in this exercise was only to construct a simple model for comparison with the 87GB results, we used the 2.7 GHz local Radio Luminosity Function (RLF) database provided by Dunlop & Peacock (1990) and made an extrapolation to our frequency. This database included the corrected Peacock (1985) data, as well as data from Toffolatti et al. (1987) and Subrahmanya & Harnett (1987). Because the time evolution of the luminosity function can be significant, this gives us only a rough estimate for the luminosity function. In practice, we fit separate power-laws to the flat- and steep-spectrum data, and then used a spectral index of -0.8 to translate the steep-spectrum power-law at 2.7 GHz to 4.85 GHz. (This was the same spectral index used by Dunlop & Peacock (1990) in translating all data to the same frequency.) For the flat-spectrum sources, a spectral index of 0 was used for the 2.7 GHz to 4.85 GHz translation, so the power-law remained unchanged. With separate flat- and steep-spectrum RLFs, we could create our selection function.

Given a population of sources, we construct the selection function $\Phi(z)$ in usual way as a fraction of sources which can be observed at given redshift. The integration of the RLFs was carried out over the redshift range $z = 0 - 5$, and over the luminosity range $P_{min} = 10^{18}$ W/Hz/sr to $P_{max} = 10^{30}$ W/Hz/sr, with $\Phi(z)$ normalized from 1 to 0 over redshifts $0 - 5$. (These limits matched those of Dunlop & Peacock's (1990) model RLF4.)

Using the $\Phi(z)$ estimate based on the Dunlop & Peacock (1990) data, we calculated the theoretical $w(\theta)$. If we take the correlation length for our radio sources to be the same as that found by Peacock & Nicholson (1991) for nearby radio galaxies $r_0 = 11h^{-1}$Mpc, then the model predicts $w(\theta)$ which is $\approx 70$ times smaller than that of the 87GB catalog. Increasing the correlation length to $r_0 = 25h^{-1}$Mpc (the correlation length of galaxy clusters) reduces the discrepancy to $\approx 15$. Because any further increase in the correlation length does not seem to be justifiable, we must look for other source of the discrepancy.

If our naive model for the luminosity function overestimates the number of radio sources at high redshifts ($z > 1$), this would significantly reduce the predicted $w(\theta)$. Reduction of the effective depth of the catalog by a factor 3–4 would bring us close to the observed correlation function. We further speculate that there could be two explanations for this: i) The number of bright radio sources declines starting at $z \approx 0.5 - 1$ as compared with the extrapolated local RLF. ii) There exists a young population of radio sources, which appeared at moderately small redshifts $z < 0.5$. Unfortunately, the uncertainties of the extrapolation of RLF from one frequency to another, extrapolations to high redshifts, and uncertainties in the evolution of the correlation function prevent us from giving any definite conclusions. It seems that our theoretical estimates fall short of the observational correlation function, but the discrepancy does not look terribly alarming when all uncertainties are taken into account.

## 5 CONCLUSIONS

The goal of this work has been to elucidate the completeness level of the 87GB survey, and to estimate $w(\theta)$ for this catalog. Additionally, some conjectures about the distribution of 87GB radio sources have also been made. The principal conclusions of our analysis can be summarized as follows.

(i) Aitoff equal area projections, surface density strips, and a periodogram analysis show that the 87GB catalog is complete to S $\geq$ 35 mJy over $20° \leq \delta < 74°$, $0^h \leq \alpha < 24^h$, and $|b| \geq 10°$. Furthermore, we caution against using sources with S < 35 mJy in statistical analyses.

(ii) The 87GB $w(\theta)$ shows evidence for the clustering of radio sources. Particularly, scales less than 2° are not inconsistent with a power-law distribution of slope -0.8. This is the *first* positive detection of an angular correlation in a large area, complete deep radio survey.

(iii) The 87GB $w(\theta)$ is significantly greater than a model correlation constructed from a $\xi(r) = (r/11h^{-1}\text{Mpc})^{-1.8}$ spatial distribution, and using the local RLF out to a redshift of $z = 5$. This might either suggest that the clustering of radio sources is stronger than the model suggests, or that the number of 87GB sources decreases at very high redshifts.

An interesting followup project would be to combine the 87GB northern sky survey with the PMN southern sky survey to achieve nearly all-sky coverage at 4.85 GHz. A comparison of this $w(\theta)$ with a more detailed model should prove very interesting. Furthermore, an even more ambitious project would be an analysis of the upcoming VLA all-sky survey at 20 cm. If this survey's sources could be coupled with the millions of images and redshifts expected from the Sloan Digital Sky Survey, the resultant 2-point analysis could yield an unprecedented investigation of radio source clustering.

This work was partially supported by NSF grants AST-9012353 and AST-9317596 to JOB. We thank Jasper Wall, Jim Condon, Ken Kellerman, Andrew Hamilton, and John Peacock for their input and suggestions on this work.

# References

Becker, R. H., White, R. L., and Edwards, A. L. 1991, ApJS, 75, 1
Bennett, A. 1962, MNRAS, 68, 165
Collins, C. A., Nichol, R. C., and Lumsden, S. L. 1992, MNRAS, 254, 295
Condon, J. J. 1992, private communication
Condon, J. J., and Broderick, J. J. 1985, AJ, 90, 2540
Condon, J. J., and Broderick, J. J. 1986, AJ, 91, 1051
Condon, J. J., Broderick, J. J., and Seielstad, G. A. 1989, AJ, 97, 1064 (CBS)
Dunlop, J. S., and Peacock, J. A. 1990, MNRAS, 247, 19
Gower, J. F. R., Scott, P. F., and Wills, D. 1967, MNRAS, 71, 49
Gregory, P. C., and Condon, J. J. 1991, ApJS, 75, 1011 (87GB)
Griffith, M. R. 1993, PhD thesis, Massachusetts Institute of Technology
Groth, E. J., and Peebles, P. J. E. 1977, AJ, 217, 385
Hamilton, A. J. S. 1993a, ApJ, in press
Hamilton, A. J. S. 1993b, private communication
Landy, S. D., and Szalay, A. S. 1993, ApJ, 412, 64
Large, M. I., Mills, B. Y., Little, A. G., Crawford, D. F., and Sutton, J. M. 1981, MNRAS, 240, 329
Limber, D. N. 1953, ApJ, 117, 134
Maddox, S.J., Efsthatiou, G., Sutherland, W. J. & Loveday, J. 1990, MNRAS, 242, 24p
Peacock, J. A. 1985, MNRAS, 217, 601
Peacock, J. A., and Nicholson, D. 1991, MNRAS, 253, 307
Peebles, P. J. E. 1980, The Large-Scale Structure of the Universe (Princeton: Princeton University Press)
Peebles, P. J. E., and Hauser, M. G. 1974, ApJS, 28, 19
Picard, A. 1991, ApJ, 368, L7
Pilkington, J. D. H., and Scott, P. F. 1965, MmRAS, 69, 183
Shaver, P. A., and Pierre, M. 1989, A&A, 220, 35
Subrahmanya, C. R., and Harnett, J. I. 1987, MNRAS, 225, 297
Toffolatti, L., Franceschini, A., De Zotti, G., and Danese, L. 1987, A&A, 184, 7
Webster, A. S. 1977, in Radio Astronomy and Cosmology, IAU Symposium No. 74, ed. D. L. Jauncey (Dordrecht: Reidel), 75

Table 1: Aitoff Region Parameters

| Figure | $\alpha$ | $\delta$ | $|b|$ | S (mJy) | number of sources |
|---|---|---|---|---|---|
| 8a | $0^h$ to $24^h$ | $20°$ to $74°$ | 10 | 35 | 21,490 |
| 8b | $0^h$ to $24^h$ | $20°$ to $74°$ | 15 | 35 | 18,839 |
| 8c | $7.5^h$ to $18.5^h$ | $20°$ to $74°$ | - | 35 | 12,577 |
| 8d | $9^h$ to $17^h$ | $20°$ to $74°$ | - | 35 | 9,183 |

Table 2: Power-law Fits to 87GB $w(\theta)$

| Fit | a | b | $\chi^2$ |
|---|---|---|---|
| solid | $3.29 \times 10^{-3}$ | -1.27 | 12.72 |
| dashed | $4.01 \times 10^{-3}$ | -0.80 | 12.88 |
| dotted | $6.32 \times 10^{-3}$ | -0.76 | 1.10 |

Figure 1: 1987 Green Bank 4.85 GHz catalog containing 54,579 sources with angular sizes $\lesssim 10\rlap{.}'5$ and S $\gtrsim$ 25 mJy.

Figure 2: 87GB catalog over the flux density intervals (a) $25 \leq S < 30$ mJy, and (b) $35 \leq S < 40$ mJy.

Figure 3: Surface number density of 1° declination strips for the S $\geq$ 35 mJy 87GB catalog.

Figure 4: Power spectrum of the 1° strip surface number density plot for the S $\geq$ 35 mJy 87GB catalog. The dashed curves represent the $2\sigma$ power fluctuations for a corresponding random sample.

Figure 5: Complete 87GB catalog of 21,490 sources, with S $\geq$ 35 mJy and over the region $20° \leq \delta < 74°$, $0^h \leq \alpha < 24^h$, and $|b| \geq 10°$.

Figure 6: $w(\theta)$ over 140° for the complete 87GB catalog. The open triangles correspond to the DD/RR formulation, the filled squares to the DD/DR formulation, and the filled circles to the (DD·RR)/(DR)² formulation. The error bar magnitude is identical for all formulations.

Figure 7: $w(\theta)$ over 5° for the complete 87GB catalog. Once again, the open triangles correspond to the DD/RR formulation, the filled squares to the DD/DR formulation, and the filled circles to the (DD·RR)/(DR)² formulation. The error bar magnitude is identical for all formulations.

Figure 8: $w(\theta)$ for the four different regions listed in Table 1.

Figure 9: $w(\theta)$ for different flux density cutoffs and for the region identical to Figure 5. (a) S $\geq$ 35 mJy. Circles show results for all sources in the region within the flux density limits. We also calculated the angular correlation function for each of original 10° strips of constant declination. Results for different strips were averaged and presented in this figure as triangles. This procedure removes any correlation due to possible mismatch between the strips. We find that within the noise limits the correlation function for the summed declination strips agrees with that found for the all-catalog correlation function. (b) S $\geq$ 45 mJy, (c) S $\geq$ 60 mJy, and (d) S $\geq$ 300 mJy.

Figure 10: "Integrated" correlation function significance $\sigma = (DD(< \theta) - RR(< \theta)) / RR^{1/2}(< \theta)$ for the 87GB catalog. The solid line corresponds to the S $\geq$ 35 mJy sample, the dashed line to the S $\geq$ 45 mJy sample, and the dotted line to the S $\geq$ 60 mJy sample.

Figure 11: Power-law fits to $w(\theta)$ of the complete 87GB catalog. correlation. All fits were from $0.3° - 1.9°$, where the solid line shows the fit to all nine data points, the dashed line the forced fit of slope -0.8, and the dotted line the fit to only the eight positive points. The filled diamond denotes the absolute value of the negative correlation at 0.9°.